\documentclass[10pt]{article}                           

\begin{document}

\def\today{Received 28~November~2002;
received in revised form 14~February~2003;
accepted 24~February~2003
\\ Communicated by Y.~Kuramoto}

\def \Oeuvres{O$\!$euvres}
\def \ie {i.e.~} 
\def\LHS{lhs~}
\def \D {\hbox{d}}
\def \Log {\mathop{\rm Log}\nolimits}
\def \sech{\mathop{\rm sech}\nolimits}
\def \Im  {\mathop{\rm Im}\nolimits}
\def \mod#1{\vert #1 \vert}
\def \jmax{J}


\title{
   Analytic solitary waves of nonintegrable equations  
\footnote{
Corresponding author RC. S2002/069. nlin.PS/0302051
}
}

\author{Micheline Musette\dag\ and Robert Conte\ddag\
{}\\
\\ \dag Dienst Theoretische Natuurkunde, Vrije Universiteit Brussel
\\ Pleinlaan 2, B--1050 Brussels, Belgium
\\                  MMusette@vub.ac.be 
{}\\
\\ \ddag Service de physique de l'\'etat condens\'e (URA 2464), CEA--Saclay
\\ F--91191 Gif-sur-Yvette Cedex, France
\\                  Conte@drecam.saclay.cea.fr
}

\maketitle

\hfill 

{\vglue -10.0 truemm}
{\vskip -10.0 truemm}

\begin{abstract} 
A major drawback of most methods to find analytic expressions for 
solitary waves is the \textit{a priori} restriction to a given class of
expressions.
To overcome this difficulty, we present a new method, 
applicable to a wide class of autonomous equations,
which builds as an intermediate information the \textit{first order}
autonomous ODE satisfied by the solitary wave.
We discuss its application to the 
cubic complex one-dimensional Ginzburg-Landau equation,
and conclude to the elliptic nature of the yet unknown 
most general solitary wave.
\end{abstract}


\noindent \textit{Keywords}:
solitary waves,
cubic complex one-dimensional Ginzburg-Landau equation,
Briot and Bouquet equations,
elliptic function,
genus,
truncation.

\noindent \textit{PACS 1995}~:
 02.30.-f   
 05.45.+b   
 42.65.-k   
 47.27.-i   

\baselineskip=12truept


\tableofcontents

\section{Introduction} 

When they are autonomous,
nonlinear partial differential equations (PDEs)
always admit 
a reduction, called traveling wave reduction,
to a nonlinear autonomous ordinary differential equation (ODE)
defined in the simplest case by $u(x,t)=U(\xi), \xi=x-ct$,
with $c$ a constant velocity.

If the PDE is integrable, for instance by the method of the 
inverse spectral transform (IST) \cite{AblowitzClarkson}, 
then there exists a whole machinery to integrate the resulting ODE
and find all the solitary waves
(maybe with some requirement of decay at infinity). 
We will not consider this case.
On the contrary, when the PDE is nonintegrable, 
the reduced ODE is also generically nonintegrable and serious
difficulties arise, which we now describe on a much studied example.

The one-dimensional cubic complex Ginzburg-Landau PDE (CGL3),
\footnote{Only the imaginary part of $\gamma$ can be absorbed in the 
definition of $A$, not the real part.
As to the contribution of a group velocity term $- i v A_x$,
it can be absorbed in the definition of $t$ since $v$ is real.}
\begin{eqnarray}
{\hskip -10.0 truemm}
& &
i A_t + p A_{xx} + q \mod{A}^2 A - i \gamma A =0,\
p q \gamma \not=0,\
(A,p,q) \in {\mathcal C},\
\gamma  \in {\mathcal R},
\label{eqCGL3}
\end{eqnarray}
which is usually written by physicists as an evolution equation 
in the case $(\Im p) (\Im q) \not=0$,
\begin{eqnarray}
{\hskip -10.0 truemm}
& &
t_0 A_t = r A + \xi_0^2 (1+ i \alpha) A_{xx} - (1 + i \beta) \mod{A}^2 A,\
(t_0,r,\xi_0,\alpha,\beta) \in {\mathcal R},
\end{eqnarray}
is a generic equation which describes many physical phenomena,
such as the propagation of a signal in an optical fiber
\cite{AgrawalBook},
spatiotemporal intermittency in spatially extended dissipative systems
\cite{MannevilleBook,vHSvS}.
Its traveling wave reduction,
\begin{eqnarray}
& &
A(x,t)=\sqrt{M(\xi)} e^{i(\displaystyle{-\omega t + \varphi(\xi)})},\
\xi=x-ct,\
(c,\omega,M,\varphi) \in {\mathcal R},
\label{eqCGL3red}
\end{eqnarray}
defines the third order system in $(M,\varphi')$,
\begin{eqnarray}
& &
\left\lbrace
\begin{array}{ll}
\displaystyle{
\frac{M''}{2 M} -\frac{{M'}^2}{4 M^2} - {\varphi'}^2
- s_i \left(\frac{c M'}{2 M} + \gamma\right)
+ s_r \left(c \varphi' + \omega\right)
+ d_r M
=0,
}
\\
\displaystyle{
\varphi'' + \varphi' \frac{M'}{M}
- s_r \left(\frac{c M'}{2 M} + \gamma\right)
- s_i \left(c \varphi' + \omega\right)
+ d_i M
=0,}
\end{array}
\right.
\label{eqCGL3ReducRealSystem}
\end{eqnarray}
with the notation for the six real parameters $d_r,d_i,s_r,s_i,g_r,g_i$,
\begin{eqnarray}
& &
d_r + i d_i = \frac{q}{p},\
s_r - i s_i = \frac{1}{p},\
g_r + i g_i = \frac{\gamma + i \omega}{p} + \frac{1}{4} c^2 s_r^2.
\end{eqnarray}

The mathematical solitary wave is the general solution,
if it exists, of the system (\ref{eqCGL3ReducRealSystem}),
equivalent to a single third order equation for $M$
(see (\ref{eqCGL3Order3}) below).

In the CGL3 case properly said $\Im (p/q) \not=0$,
the system (\ref{eqCGL3ReducRealSystem}) is only partially integrable,
which means that the general solution 
(which should depend on three arbitrary integration constants)
does not exist,
and the question is to find, in closed form,
the solution which depends on the largest possible number of arbitrary
constants,
let us call it the \textit{general analytic solution}.
This number of arbitrary constants, smaller than three,
can be computed either from singularity analysis \cite{CT1989,CM2000b},
or from topological arguments \cite{vSH},
and it is equal to one,
namely the origin $\xi_0$ of $\xi$. 
The question is therefore to find the unique particular solution of
(\ref{eqCGL3ReducRealSystem}),
\begin{eqnarray}
& &
M = f(\xi-\xi_0),
\label{eqCGL3FormalSolution}
\end{eqnarray}
which has codimension zero
(the codimension is defined as the number of constraints among 
the parameters $d_r,d_i,s_r,s_i,\gamma,c,\omega$
of the system (\ref{eqCGL3ReducRealSystem})).

At present time,
all existing methods 
(mainly the so called truncation methods,
see e.g.~the summer school lecture notes \cite{Cargese1996Musette})
have failed to find this codimension-zero solution,
they have only found several solutions with a nonzero codimension
in which $M$ is a polynomial in a variable which satisfies 
a Riccati equation. 
What we present here is not just an additional method
which could maybe yield a new particular solution
with a nonzero codimension,
but an entirely new approach which for sure will find,
sooner or later, 
the desired expression $f(\xi-\xi_0)$ in (\ref{eqCGL3FormalSolution}).
Rather than looking for an explicit, closed form expression,
we look for the first order ODE 
resulting from the elimination of $\xi_0$ between 
(\ref{eqCGL3FormalSolution}) and its derivative.
Because of the invariance by translation of $\xi$,
eliminating $\xi_0$ is equivalent to eliminate $\xi-\xi_0$,
therefore this first order ODE is autonomous.
There exist classical results of complex analysis,
mainly due to Briot and Bouquet \cite[pages 58--59]{PaiLecons},
which establish the following important points:
\begin{enumerate}
\item
the knowledge of the above mentioned nonlinear first order autonomous ODE
implies the knowledge of the expression $f(\xi-\xi_0)$,

\item
the nature of this solution $f(\xi-\xi_0)$
can only be elliptic
(i.e.~meromorphic doubly periodic in the complex plane)
or a degeneracy of elliptic (i.e. trigonometric (simply periodic)
or rational).

\end{enumerate}

Therefore,
if one is able to directly build this first order ODE,
this suppresses the need for making a good guess for the class of expressions
$f(\xi-\xi_0)$,
since this class is now an output, not an input, of the method.

The new method presented here 
is particularly suited to nonintegrable equations,
it is algorithmic, 
and is very easy to implement since it only requires the resolution
of a \textit{linear} system.

The paper is organized as follows.

In section \ref{sectionAlready_known_particular_solutions},
we recall the already known particular solutions of
(\ref{eqCGL3Order3}) in the CGL3 case,
together with their nonzero codimension.
In section \ref{sectionCorresponding_first_order_subequations},
we give the corresponding first order equations which they satisfy.
In section \ref{sectionclassical_results_on_first_order_equations},
we briefly recall the previous methods,
then
we summarize the relevant classical results.
In section \ref{sectionLinear_method},
we present the new method and its advantages.
In section \ref{sectionResultsCGL3},
we give the results for CGL3.
Finally, in section \ref{sectionDiscussion},
we indicate the class of dynamical systems to which the method 
can be applied.

\section {Particular solutions with a nonzero codimension}
\label{sectionAlready_known_particular_solutions}

Rather than the system of two coupled ODEs (\ref{eqCGL3ReducRealSystem}),
it is necessary, in order to apply our method,
to consider a single ODE,
e.g.~for the field $M=\mod{A}^2$,
obtained by eliminating $\varphi$ between 
the system (\ref{eqCGL3ReducRealSystem}),
\begin{eqnarray}
& &
\varphi' = \frac{c s_r}{2} + \frac{G'-2 c s_i G}{2 M^2( g_r - d_i M)},\
\left(\varphi' - \frac{c s_r}{2}\right)^2=\frac{G}{M^2},
\label{eqCGL3Phiprime}
\\
& &
(G'-2 c s_i G)^2 - 4 G M^2  (d_i M - g_r)^2=0,\
\label{eqCGL3Order3}
\\
& &
G=\frac{1}{2} M M'' - \frac{1}{4} M'^2
  -\frac{c s_i}{2} M M' + d_r M^3 + g_i M^2.
\end{eqnarray}

In the CGL3 case properly said $\Im(p/q)\not=0$,
only three solutions are currently known.
In all of them $M$ is a second degree polynomial in 
$ \tau=(k/2) \tanh k \xi/2$, with $k^2$ a real constant.
These are 
\begin{enumerate}

\item
a codimension-one source or propagating hole \cite{BN1985}
(heteroclinic orbit),

\item
a codimension-two pulse or solitary wave \cite{PS1977}
(homoclinic orbit),

\item
a codimension-two front or shock \cite{NB1984}
(heteroclinic orbit).

\end{enumerate}

To avoid carrying heavy expressions, 
let us make the following nonrestrictive simplification.
Out of the five 
parameters $d_r,d_i,g_r,g_i,c s_i$
of the ODE (\ref{eqCGL3Order3}),
only three are essential 
($g_r,g_i,c$, equivalent to $\gamma,\omega,c$).
Indeed, just like for NLS, $p$ and $q$ (i.e. $d_r+i d_i$ and $s_r - i s_i$)
can be rescaled to convenient numerical values,
making rational all the coefficients of the above solutions, 
such as
\begin{eqnarray}
{\hskip -10.0 truemm}
& &
p=-1-3 i,\
q= 4-3 i,\
\nonumber
\\
& &
d_r= \frac{1}{2},\
d_i= \frac{3}{2},\
s_r=-\frac{1}{10},\
s_i=-\frac{3}{10}.
\label{eqCGL3Num}
\end{eqnarray}

Each of the three solutions depends on an additional sign
and, for these numerical values, the two sets of three solutions are

\begin{eqnarray}
& &
\left\lbrace
\matrix{
\displaystyle{
M= -2 \left\lbrack\left(\tau-\frac{c}{20}\right)^2+\left(\frac{c}{10}\right)^2
      \right\rbrack,\
\varphi'-\frac{c s_r}{2}=
\tau + \frac{c}{20} + \frac{c}{5 M} \left(\tau^2 - \frac{k^2}{4}\right),\
}
\hfill \cr
\displaystyle{
k^2=- 7 \left(\frac{c}{10}\right)^2 - \frac{4}{3} g_r,\
3 g_i + 2 g_r + \frac{3 c^2}{50}=0,\
}
\hfill \cr}
\right.
\label{eqCGL3Hole-}
\\
& &
\left\lbrace
\matrix{
\displaystyle{
M= -2 \left(\tau^2 - \frac{k^2}{4}\right),\
\varphi'-\frac{c s_r}{2}=\tau,
}
\hfill \cr
\displaystyle{
k^2=2 g_r,\
c=0,\ g_i=0,\
}
\hfill \cr}
\right.
\label{eqCGL3Pulse-}
\\
& &
\left\lbrace
\matrix{
\displaystyle{
M= -2 \left(\tau \pm \frac{k}{2}\right)^2,\
\varphi'-\frac{c s_r}{2}= \tau - \frac{c}{20},
}
\hfill \cr
\displaystyle{
k^2=\left(\frac{c}{10}\right)^2,\
g_r=0,\ g_i - \frac{c^2}{50}=0,\
}
\hfill \cr}
\right.
\label{eqCGL3Front-}
\end{eqnarray}
and
\begin{eqnarray}
& &
\left\lbrace
\matrix{
\displaystyle{
M= 4 \left\lbrack\left(\tau - \frac{c}{20}\right)^2
  + \left(\frac{c}{20}\right)^2
      \right\rbrack,\
\varphi'-\frac{c s_r}{2}=
-2 \tau + \frac{c}{20} + \frac{c}{5 M} \left(\tau^2 - \frac{k^2}{4}\right),\
}
\hfill \cr
\displaystyle{
k^2=- \left(\frac{c}{10}\right)^2 + \frac{2}{3} g_r,\
3 g_i - g_r + \frac{3 c^2}{80}=0,\
}
\hfill \cr}
\right.
\label{eqCGL3Hole+}
\\
& &
\left\lbrace
\matrix{
\displaystyle{
M= 4 \tau^2,\
\varphi'-\frac{c s_r}{2}=-2 \tau,
}
\hfill \cr
\displaystyle{
k^2=\frac{2}{3} g_r,\
c=0,\ 3 g_i - g_r =0,\
}
\hfill \cr}
\right.
\label{eqCGL3Pulse+}
\\
& &
\left\lbrace
\matrix{
\displaystyle{
M= 4 \left(\tau \pm \frac{k}{2}\right)^2,\
\varphi'-\frac{c s_r}{2}=- 2 \tau - \frac{c}{10},
}
\hfill \cr
\displaystyle{
k^2=\left(\frac{c}{10}\right)^2,\
g_r=0,\ g_i - \frac{c^2}{50}=0.
}
\hfill \cr}
\right.
\label{eqCGL3Front+}
\end{eqnarray}

\textit{Remark}.
It has been predicted \cite{vanHecke}
the existence of a fourth physically interesting solution,
which is a codimension-one 
homoclinic hole solution with an arbitrary velocity $c$.
{}From the study of the structure of singularities of the CGL3 equation
\cite{CM1993},
the most likely analytic expression to represent it
would be a polynomial in both $\tanh k \xi/2$ and $\sech k \xi/2$
of global degree two,
\begin{eqnarray}
& &
M=\left(\frac{3 \sqrt{9 d_r^2 + 8 d_i^2}}{2 d_i^2} \tanh + c_1\right) \sech 
+  \frac{9 d_r}{2 d_i^2}  \tanh^2 + c_3 \tanh + c_4,
\label{eqCGL3TwoFamiliesSolution}
\end{eqnarray}
with the homoclinic condition $c_3=0$.
We have checked that, unfortunately,
such a solution does not exist in the CGL3 case $\Im(p/q)\not=0$.

\section {The corresponding first order subequations}
\label{sectionCorresponding_first_order_subequations}

The only arbitrary constant in the two sets of three solutions of CGL3
is the origin $\xi_0$ of $\xi$,
which represents the location of the movable pole 
(movable means which depends on the arbitrary initial conditions). 
By eliminating it between the expressions for $M$ and $M'$,
one obtains first order ODEs.
For the set (\ref{eqCGL3Hole-})--(\ref{eqCGL3Front-}), these are
\begin{eqnarray}
& &
\left(M' + \frac{c}{5} M + \frac{c^3}{250}\right)^2 
+ 2 \left(M + \frac{c^2}{50}\right)
    \left(M - \frac{c^2}{50} - \frac{2}{3} g_r\right)^2 =0,\
\label{eqsubeqCGL3Hole-}
\\
& &
{M'}^2 + 2 (M-g_r) M^2=0,
\label{eqsubeqCGL3Pulse-}
\\
& &
\left(M' + \frac{c}{5} M\right)^2 + 2 M^3=0,
\label{eqsubeqCGL3Front-}
\end{eqnarray}
with the respective constraints on $(g_r,g_i,c)$ already indicated,
and similarly for the set (\ref{eqCGL3Hole+})--(\ref{eqCGL3Front+}), 
\begin{eqnarray}
& &
\left(M' + \frac{c}{5} M - \frac{c^3}{500}\right)^2 
-   \left(M - \frac{c^2}{100}\right)
    \left(M + \frac{c^2}{100} - \frac{2}{3} g_r\right)^2 =0,\
\label{eqsubeqCGL3Hole+}
\\
& &
{M'}^2 - M \left(M -\frac{2}{3} g_r\right)^2=0,
\\
& &
\left(M' + \frac{c}{5} M\right)^2 - M^3=0.
\label{eqsubeqCGL3Front+}
\end{eqnarray}

These six first order ODEs will be called
\textit{subequations} of the third order ODE (\ref{eqCGL3Order3}),
since (\ref{eqCGL3Order3}) is a differential consequence
of any of the six first order ODEs,
with the respective constraints on $(g_r,g_i,c)$.

Another information,
quite useful to get some hint on 
the sought expression $f(\xi-\xi_0)$ in (\ref{eqCGL3FormalSolution}),
is made of the similar expressions
for the unique integrable case of the CGL3 PDE (\ref{eqCGL3}),
namely $q/p \in {\mathcal R}, \gamma=0$,
which defines the nonlinear Schr\"odinger equation (NLS),
\begin{eqnarray}
& &
i A_t + p A_{xx} + q \mod{A}^2 A =0,\
p q \not=0,\
A \in {\mathcal C},\
(p,q)  \in {\mathcal R}.
\label{eqNLS}
\end{eqnarray}
In this case,
as opposed to the generic CGL3 case,
the general solution of the system (\ref{eqCGL3ReducRealSystem}) exists.
Indeed, this system admits two first integrals $K_1,K_2$,
\begin{eqnarray}
& &
M \varphi' - \frac{c}{2 p} M=K_1.
\label{eqNLSFirstIntegral1}
\\
& &
\frac{{M'}^2}{M} +\frac{2 q}{p} M^2 
+\left(\frac{c^2 - 4 \omega p}{p^2}\right) M + 4 \frac{K_1^2}{M}=K_2,
\label{eqNLSFirstIntegral2}
\end{eqnarray}
so the general solution is the singlevalued closed form expression 
\begin{eqnarray}
{\hskip -9.0truemm}
& &
M=-2 \frac{p}{q} \left(\wp(\xi-\xi_0,g_2,g_3) - e_0\right),\
\varphi' - \frac{c}{2 p} = \frac{K_1}{M},
\label{eqNLSmod2}
\\
{\hskip -9.0truemm}
& & 
e_0=\frac{4 \omega p - c^2}{12 p^2},\
K_1^2=- \left(\frac{p}{q}\right)^2 (4 e_0^3-g_2 e_0-g_3),\
K_2=-4 \frac{p}{q} \left(6 e_0^2 - \frac{g_2}{2}\right),
\end{eqnarray}
in which $\wp$ is the Weierstrass elliptic function defined by
\begin{eqnarray}
& &
{\wp'}^2=4 \wp^3 - g_2 \wp - g_3.
\end{eqnarray}
It depends on four fixed constants,
$p,q,c,\omega$ (fixed means which appears in the definition of the ODE),
and three movable constants.
Out of these three,
one ($\xi_0$) just represents the translational invariance,
and the two important ones are $(g_2,g_3)$ or equivalently $(K_1,K_2)$.

The two physically meaningful solitary waves of NLS
(the bright one and the dark one)
are obtained from this mathematical solitary wave
by imposing some decaying conditions
at infinity on the real axis of the complex variable $\xi$.
The same feature will certainly apply after the sought solution 
of CGL3 has been found.

\section {The classical results on first order equations}
\label{sectionclassical_results_on_first_order_equations}

The six subequations of CGL3 all have the same degree two 
(in the highest derivative $M'$),
just like the (unique) similar subequation (\ref{eqNLSFirstIntegral2})
for NLS,
but the difference with NLS is the absence of first integrals $K_1,K_2$.

The six subequations are by construction in a one-to-one correspondence with
the six solutions of 
section \ref{sectionAlready_known_particular_solutions},
therefore it is equivalent, in principle,
to search for subequations or for solutions.
However, there is a big difference between these two classes of methods.

In every method which directly searches for solutions,
one must \textit{a priori} assume a given class of possible solutions,
then compute whether there indeed is some solution in this class.
All these methods can therefore be qualified as ``sufficient''. 
The main existing such methods are
\begin{enumerate}

\item
the one-family truncation method \cite{WTC,CM1989}
which assumes the class of polynomials in $\tanh$, here
\begin{eqnarray}
& & M=c_2 \tau^2 + c_1 \tau + c_0,\ c_2 \not=0,\ 
\tau=(k/2) \tanh k \xi/2,
\end{eqnarray}

\item
the two-family truncation method \cite[App.~A]{CM1993} \cite{MC1994},
which assumes the class of polynomials in $\tanh$ and $\sech$, here
\begin{eqnarray}
& & M=(d_1 \tau + d_0) \sigma + c_2 \tau^2 + c_1 \tau + c_0,\
(d_1,c_2) \not=(0,0),\
\\
& &
  \tau=(k/2) \tanh k \xi/2,\
\sigma=(k/2) \sech k \xi/2,\
\end{eqnarray}

\item
the elliptic method \cite{KudryashovElliptic,Samsonov1994}, 
which assumes the class of polynomials in the Weierstrass function
$\wp(\xi)$ and its derivative $\wp'(\xi)$, here
\begin{eqnarray}
& & M=c_1 \wp + c_0,\ c_1 \not=0.
\end{eqnarray}

\end{enumerate}

Therefore it is easy to build an ODE which escapes these methods,
e.g.~any differential consequence of the first order ODE
\begin{eqnarray}
& & 
{M'}^2+\left(12 M^2 - \frac{3}{2}\right) M'
 + 36 M^4 - \frac{17}{2} M^2 + \frac{1}{2}=0,
\end{eqnarray}
whose solution is 
\begin{eqnarray}
& & 
M=\frac{\tanh(\xi-\xi_0)}{2+\tanh^2(\xi-\xi_0)}.
\end{eqnarray}

On the contrary,
the search for first order autonomous subequations
requires no \textit{a priori} assumption at all,
and, from classical results which we now recall, 
the knowledge of the first order subequation is indeed equivalent
to the knowledge of the explicit expression (\ref{eqCGL3FormalSolution}).
As opposed to the previous methods, 
which are ``sufficient'' as said above, 
the proposed method can be qualified as ``necessary''.

Consider the class of first order nonlinear algebraic ODEs,
\begin{eqnarray}
& &
F(M,M') =0,
\label{eqOrder1Autonomous}
\end{eqnarray}
in which $F$ is a polynomial of two variables.
The unknown first order ODE 
resulting from the elimination of $\xi_0$ between 
(\ref{eqCGL3FormalSolution}) and its derivative,
under the (reasonable) assumption that it is algebraic,
belongs to this class.

A first theorem was established by Briot and Bouquet
(see, e.g., \cite[Vol.~II, \S 139 p.~284]{Valiron}):
\textit{If the general solution of the autonomous ODE 
(\ref{eqOrder1Autonomous}) is singlevalued,
then this solution is
either an elliptic function,
or a rational function of $e^{a x}$, $a$ being some constant, 
or a rational function of $x$.
}
Note that the third case is a degeneracy of the second one, 
itself a degeneracy of the first one.

A second result,
of immediate practical use,
is due to Painlev\'e \cite[pages 58--59]{PaiLecons}.
\textit{If the general solution of the autonomous ODE 
(\ref{eqOrder1Autonomous}) is singlevalued,
then the necessary form of this ODE is
\begin{eqnarray}
& &
F(M,M') \equiv
 \sum_{k=0}^{m} \sum_{j=0}^{2m-2k} a_{j,k} M^j {M'}^k=0,\ a_{0,m}=1,
\label{eqsubeqODEOrderOnePP}
\end{eqnarray}
in which $m$ is a positive integer and the ${a_{j,k}}'s$ are constants.}

We now present a method which, given the integer $m$,
directly computes the coefficients of (\ref{eqsubeqODEOrderOnePP}).

\section{The linear method based on the Laurent series}
\label{sectionLinear_method}

The general analytic solution of the ODE (\ref{eqCGL3Order3}),
which depends on the single movable constant $\xi_0$,
is known locally,
this is the Laurent series \cite{CT1989,CM1993}
\begin{eqnarray}
& & 
M_\pm = \frac{9 d_r \pm 3 \Delta}{2 d_i^2} \chi^{-2}
\left(1 + \frac{c s_i}{3} \chi + {\mathcal O}(\chi^2)\right),\
\chi = \xi - \xi_0,
\label{eqCGL3mod2Laurent}
\end{eqnarray}
in which 
$\xi_0$ is the location of a movable double pole, and 
$\Delta$ is the positive fixed constant
\begin{eqnarray}
& &
\Delta=\sqrt{9 d_r^2+8 d_i^2}.
\end{eqnarray}

If one could eliminate $\xi_0$, i.e.~$\xi-\xi_0$,
between the Laurent series and its derivative,
this would settle the question of establishing 
the codimension-zero first order subequation.

The only piece of information we have on 
the unknown global expression $f(\xi-\xi_0)$ 
is its local representation by this Laurent series.
In particular, there is no mathematical proof yet that $f(\xi-\xi_0)$ 
is globally singlevalued.
However, a numerical investigation \cite{YCM2003} 
of the singularities of $f$ in the complex plane
displays a pattern of singularities
which looks like a doubly periodic set of poles and zeros,
i.e. the signature of some elliptic function.
Let us therefore assume the singlevaluedness of $f$.

The algorithm to convert this local information (the Laurent series)
into a global one (the codimension-zero first order subequation)
is the following.

\begin{enumerate}

\item
Choose a positive integer $m$ and define the Briot and Bouquet
first order ODE (\ref{eqsubeqODEOrderOnePP}).
It contains $(m+1)^2-1$ unknown constants $a_{j,k}$.

\item
Compute $\jmax$ terms of the Laurent series,
   with $\jmax$ slightly greater than $(m+1)^2-1$,
\begin{eqnarray}
& &
M=\chi^p \left(\sum_{j=0}^{\jmax} M_j \chi^j+{\mathcal O}(\chi^{j+1})\right),
\label{eqLaurent}
\end{eqnarray}
with in the CGL3 case $p=-2$.

\item
Require the Laurent series to satisfy the Briot and Bouquet ODE,
i.e.~require the identical vanishing of the Laurent series for the \LHS\
$F(M,M')$ up to the order $\jmax$
\begin{eqnarray}
& &
F \equiv \chi^{D} \left(\sum_{j=0}^{\jmax} F_j \chi^j
 + {\mathcal O}(\chi^{\jmax+1})
\right),\
D=m(p-1),\
\forall j\ : \ F_j=0.
\end{eqnarray}
If it has no solution for $a_{j,k}$, increase $m$ and return to first step.

\item
For every solution,
integrate the first order autonomous ODE (\ref{eqsubeqODEOrderOnePP}).

\end{enumerate}

The core of the method is the third step, 
in which the system of $\jmax+1$ equations $F_j=0$ 
in the $(m+1)^2-1$ unknowns $a_{j,k}$
is \textit{linear} and \textit{overdetermined},
therefore quite easy to solve.
Indeed, it has generically no solution until $m$ reaches its correct value.

As to the fourth step, it is also quite easy.
Indeed, 
when the first order autonomous ODE (\ref{eqsubeqODEOrderOnePP})
has the property that its general solution is singlevalued,
there exists a classical method due to Poincar\'e to perform the 
integration,
and it has been implemented
at least in one computer algebra language \cite{MapleAlgcurves}.
One first has to compute 
the genus of the algebraic curve $F(M,M')=0$,
which can only be zero or one because of the singlevaluedness.
If the genus is zero,
the general solution of (\ref{eqsubeqODEOrderOnePP})
is a rational function of $e^{a \xi}$, with $a$ constant,
or a rational function of $\xi$. 
If the genus is one,
the general solution is an elliptic function of $\xi$,
which can always be put in the canonical form
of a rational function of $\wp(\xi)$ and $\wp'(\xi)$.

The above method clearly includes the three previous ones mentioned
in Section \ref{sectionCorresponding_first_order_subequations},
which are recovered with the respective values 
$m=-p,-2p,-2p$.
Therefore it can only find more results,
and more easily since the main computation (third step) is just linear.

\textit{Remark 1}.
In addition to the Briot and Bouquet selection rule $j+2k \le 2 m$,
one also has the selection rule $m(p-1) \le j p + k (p-1)$,
which, when $p\not=-1$,
further restricts the number of coefficients $a_{j,k}$.

\textit{Remark 2}.
There is no upper bound on $m$,
and this is the only inconvenient of the method.
However, there exists a lower bound,
obtained by the requirement that the most singular term 
${M'}^m \sim M_0^m \chi^{m (p-1)}$
must balance some other term in (\ref{eqsubeqODEOrderOnePP}).
For $p=-2$ (the CGL3 or NLS case),
this forbids $m=1$ (because $M'$ cannot balance $M^2$, nor $M$, nor $1$),
and the loop must start at $m=2$.

\section{Results for CGL3}
\label{sectionResultsCGL3}

With the numerical values (\ref{eqCGL3Num}), one has $\Delta=9/2$,
and the two Laurent series are
\begin{eqnarray}
& &
M_- = \chi^{-2} \left(
 -2+\frac{c}{5} \chi
 +\left(\frac{g_r}{3} -\frac{g_i}{6} -\frac{c^2}{200}\right) \chi^2
 + {\mathcal O}(\chi^3)\right),
\label{eqLaurentCGL3-}
\\
& &
M_+ = \chi^{-2} \left(
 4-\frac{2c}{5} \chi
 +\left(\frac{16g_r}{39} +\frac{4 g_i}{39} +\frac{19 c^2}{1300}\right) \chi^2
 + {\mathcal O}(\chi^3)\right).
\label{eqLaurentCGL3+}
\end{eqnarray}

The existence of two Laurent series, rather than just one,
is a feature which the subequation must also possess,
and this has the effect of setting the lower bound to $m=4$ instead of $2$.
Indeed, 
the lowest degree subequations
\begin{eqnarray}
F_2 & \equiv & {M'}^2
 + {M'} (a_{1,1} M   + a_{0,1})
       + a_{3,0} M^3 + a_{2,0} M^2 + a_{1,0} M + a_{0,0}=0,
\label{eqsubeqODEOrderOnep2m2}
\\
F_3 & \equiv & {M'}^3 
+ {M'}^2 (a_{1,2} M   + a_{0,2})
+ {M'}   (a_{3,1} M^3 + a_{2,1} M^2 + a_{1,1} M + a_{0,1})
\nonumber
\\
& &
+         a_{4,0} M^4 + a_{3,0} M^3 + a_{2,0} M^2 + a_{1,0} M + a_{0,0}
=0,
\label{eqsubeqODEOrderOnep2m3}
\end{eqnarray}
have the respective dominant terms
${M'}^2 + a_{3,0} M^3$ and ${M'}^3 + a_{3,1} {M'} M^3$,
which define only one family of movable double poles.

To give a detailed account of the method,
let us nevertheless start with the lowest bound $m=2$,
for which (\ref{eqsubeqODEOrderOnep2m2}) can only be satisfied
by one series, e.g.~(\ref{eqLaurentCGL3-}),
thus preventing the full desired result to be obtained.
The six coefficients $a_{j,k}$ of (\ref{eqsubeqODEOrderOnep2m2})
are first computed as the unique solution of the linear system of
six equations $F_j=0,j=0,1,2,3,4,6$.
Then the $\jmax+1-6$ remaining equations $F_j=0,j=5,7:\jmax$,
which only depend on the fixed parameters $(g_r,g_i,c)$,
have the greatest common divisor (gcd) $3 g_i + 2 g_r + 3 c^2/50$,
and this factor defines the first solution (\ref{eqCGL3Hole-}),
namely the codimension-one propagating hole of Bekki and Nozaki.
After division par this gcd,
the system of three equations $F_j=0,j=5,7,8$,
provides two and only two other solutions, 
which are (\ref{eqsubeqCGL3Pulse-}) and (\ref{eqsubeqCGL3Front-}),
with the respective constraints $(c=0,g_i=0)$ and $(g_r=0,50 g_i-c^2=0)$,
and all the remaining equations $F_j=0,j \ge 9$, are identically satisfied.

Therefore, with this lower bound $m=2$,
one already recovers all the presently known first order subequations.
Finally, for each of the three subequations,
the fourth step finds a zero value for the genus
and returns the general solution as a rational function of 
$e^{a (\xi-\xi_0)}$,
which basic trigonometric identities then allow to convert to the
second degree polynomials in $(k/2) \tanh k (\xi-\xi_0)/2$
listed in (\ref{eqCGL3Hole-})--(\ref{eqCGL3Front-}).

With the correct two-family lower bound $m=4$,
which corresponds to $18$ unknowns $a_{j,k}$ and 
at least $24$ terms in the series,
we have checked that there is no solution other than the above three.
This situation is quite similar to the absence of solution in the class
(\ref{eqCGL3TwoFamiliesSolution}),
and it just reflects the difficulty of the CGL3 equation.

The case $m=8$ 
($60$ unknowns $a_{j,k}$ and at least $66$ terms in the series)
is currently under investigation
but preliminary results seem to indicate the absence of any new solution,
and we are now automatizing the computer algebra program
in order to handle much larger values of $m$.
Indeed,
if one admits the singlevaluedness of 
$f(\xi-\xi_0)$ in (\ref{eqCGL3FormalSolution}),
then for sure there exists some positive integer $m_0$,
alas without upper bound,
at which the result will be obtained.

\section{Domain of applicability of the method}
\label{sectionDiscussion}

As we have seen,
the present method contains the three main methods already available,
and its cost is minimal since the main step is a linear computation.

The two key assumptions behind this ``subequation method'' are,
\begin{enumerate}

\item
a Laurent series should exist,

\item
a first order autonomous algebraic subequation should exist.

\end{enumerate}

Its best applicability is therefore nonintegrable
$N$-th order autonomous nonlinear ODEs
admitting a Laurent series which only depends on one movable constant,
such as the present CGL3 ODE (\ref{eqCGL3Order3})
or the traveling wave reduction of the Kuramoto-Sivashinsky equation
\cite{CM1989,YCM2003}. 

Two examples of inapplicability are
\begin{enumerate}

\item
the Lorenz model,
in which the Laurent series generically does not exist
and has to be replaced by a psi-series \cite{Segur},

\item
the autonomous ODE $M'''-12 M M' -1=0$,
which admits the first Painlev\'e transcendent as its general solution,
a case in which no first order subequation exists.

\end{enumerate}

\section{Conclusion}

The advantage of the Laurent series is to exclude the contribution of
chaos in a dynamical system,
its disadvantage is to be a local piece of information.
The present method converts this local information into the
global one of a first order autonomous subequation,
\textit{without any other assumption} than the singlevaluedness
of the general solution of this first order ODE.
This is therefore a significant improvement over previous methods,
which all require an additional assumption restricting
the sought solutions to some class of analytic expressions.

{}From classical results,
and again under the single assumption of singlevaluedness,
we conclude that the yet unknown codimension-zero solution is elliptic,
i.e. equal to some rational function of 
$\wp(\xi-\xi_0)$ and $\wp'(\xi-\xi_0)$.

\section*{Acknowledgments}

The authors acknowledge the financial support of the Tournesol grant 
T2003.09 and the support of CEA.


\vfill \eject

\begin{thebibliography}{99}

\bibitem {AblowitzClarkson}
M.~J.~Ablowitz and P.~A.~Clarkson,
\textit{Solitons, nonlinear evolution equations and inverse scattering}
(Cambridge University Press, Cambridge, 1991).

\bibitem{AgrawalBook} G.~P.~Agrawal,
\textit{Nonlinear fiber optics}
(Academic press, Boston, 1989).

\bibitem{BN1985} N.~Bekki and K.~Nozaki,
Formations of spatial patterns and holes in the generalized 
Ginzburg-Landau equation,
Phys.~Lett.~A {\bf 110} (1985) 133--135.

\bibitem{CT1989} F.~Cariello and M.~Tabor,
Painlev\'e expansions for nonintegrable evolution equations,
Physica D {\bf 39} (1989) 77--94.

\bibitem{CM1989} R.~Conte and M.~Musette,
Painlev\'e analysis and B\"acklund transformation in the 
Kuramoto-Sivashinsky equation,
J.~Phys.~A {\bf 22} (1989) 169--177.

\bibitem{CM1993} R.~Conte and M.~Musette,
Linearity inside nonlinearity:
exact solutions to the complex Ginzburg-Landau equation,
Physica D {\bf 69} (1993) 1--17.
 
\bibitem{CM2000b} R.~Conte and M.~Musette,
Analytic expressions of hydrothermal waves,
Reports on mathematical physics {\bf 46} (2000) 77--88.
nlin.SI/0009022 

\bibitem{vanHecke} M.~van Hecke,                              
Building blocks of spatiotemporal intermittency,
Phys.~Rev.~Lett.~{\bf 80} (1998) 1896--1899.

\bibitem{vHSvS} M.~van Hecke, C.~Storm, and W.~van Saarlos,   
Sources, sinks and wavenumber selection in coupled CGL equations and
experimental implications for counter-propagating wave systems,
Physica D  {\bf 133} (1999) 1--47. Patt--sol/9902005.
 
\bibitem{KudryashovElliptic} N.~A.~Kudryashov,
Exact solutions of a generalized equation of Ginzburg-Landau,
Matematicheskoye modelirovanie {\bf 1} (1989) 151--158.
[English~: none?].

\bibitem{MapleAlgcurves} Mark van Hoeij,
package ``algcurves'', Maple V (1997).
\verb+http://www.math.fsu.edu/~hoeij/algcurves.html +

\bibitem{MannevilleBook} P.~Manneville,
\textit{Dissipative structures and weak turbulence}
(Academic Press, Boston, 1990).
French adaptation:
\textit{Structures dissipatives, chaos et turbulence}
(Al\'ea-Saclay, Gif-sur-Yvette, 1991).

\bibitem{Cargese1996Musette} M.~Musette,
Painlev\'e analysis for nonlinear partial differential equations,
{\it The Painlev\'e property, one century later}, 
517--572,
ed.~R.~Conte,
CRM series in mathematical physics (Springer, New York, 1999).

\bibitem{MC1994} M.~Musette and R.~Conte,
The two--singular manifold method, I. Modified KdV and sine-Gordon equations,
J.~Phys.~A {\bf 27} (1994) 3895--3913.

\bibitem{NB1984} K.~Nozaki and N.~Bekki,
Exact solutions of the generalized Ginzburg-Landau equation,
J.~Phys.~Soc.~Japan {\bf 53} (1984) 1581--1582.

\bibitem{PaiLecons} P.~Painlev\'e,
{\it Le\c{c}ons sur la th\'eorie analytique des \'equations diff\'erentielles}
(Le\c{c}ons de Stockholm, 1895)
(Hermann, Paris, 1897).
Reprinted, {\it \Oeuvres\ de Paul Painlev\'e}, vol.~I
(\'Editions du CNRS, Paris, 1973).

\bibitem{PS1977} N.~R.~Pereira and L.~Stenflo,
Nonlinear Schr\"odinger equation including growth and damping,
Phys.~Fluids {\bf 20} (1977) 1733--1743.
 
\bibitem{Samsonov1994} A.~M.~Samsonov,             
Nonlinear strain waves in elastic waveguides,
{\it Nonlinear waves in solids}, 349--382
eds.~A.~Jeffrey and J.~Engelbrecht
(Springer-Verlag, Wien, 1994).

\bibitem{vSH} W.~van Saarloos and P.~C.~Hohenberg,
Fronts, pulses, sources and sinks in generalized complex 
Ginzburg-Landau equations,
Physica D {\bf 56} (1992) 303--367.
Erratum {\bf 69} (1993) 209.

\bibitem{Segur} H.~Segur, 
Solitons and the inverse scattering transform,
{\it Topics in ocean physics},
235--277,
eds.~A.~R.~Osborne and P.~Malanotte Rizzoli
(North-Holland publishing co., Amsterdam, 1982).

\bibitem{Valiron} G.~Valiron,                              
\textit{Cours d'analyse math\'ematique},
2i\`eme \'ed. (Masson, Paris, 1950).
 
\bibitem{WTC} J.~Weiss, M.~Tabor, and G.~Carnevale,
The Painlev\'e property for partial differential equations,
J.~Math.~Phys.~{\bf 24} (1983) 522--526.
 
\bibitem{YCM2003} Yee T.-l., R.~Conte, and M.~Musette,
Sur la ``solution analytique g\'en\'erale'' d'une \'equation
   diff\'erentielle chaotique du troisi\`eme ordre,
13 pages,
IRMA Lectures in  Mathematics and Theoretical Physics {\bf }
(de Gruyter, Berlin, 2003),
accepted for publication.
Preprint S2002/070.
nlin.PS/0302nnn.
Journ\'ees de calcul formel, Strasbourg, IRMA, 21--22 mars 2002.

\end{thebibliography}
\end{document}